
\magnification=1200
\parskip=\medskipamount \overfullrule=0pt
\parindent=20truept
\font \titlefont=cmr10 scaled \magstep3
\font \namefont=cmr10 scaled \magstep1

\def\singlespace{\baselineskip=\normalbaselineskip}

\newcount\firstpageno \firstpageno=2
\footline={\ifnum\pageno<\firstpageno{\hfil}\else{\hfil
                                                  \rm\folio\hfil}\fi}
\def\frac#1/#2{{\textstyle{#1\over #2}}}
\def\degrees{\hbox{${}^\circ$\hskip-3pt .}}
\def\pp{\par\hangindent=.125truein \hangafter=1}
\def\aref#1;#2;#3;#4{\pp #1, {\it #2}, {\bf #3}, #4}
\def\abook#1;#2;#3{\pp #1, {\it #2}, #3}
\def\arep#1;#2;#3{\pp #1, #2, #3}
\def\T{\widetilde{T}}

\def\nhat{{\bf n}}
\def\n{{\bf n}}

\def\nhatav{{\bf \overline{n}}}

\def\cd{\!\cdot\!}

\def\Wl{W_{\ell}}
\def\Pl{P_{\ell}}
\def\half{{\textstyle{1\over2}}}
\def\third{{\textstyle{1\over3}}}
\def\gtwid{\mathrel{\raise.3ex\hbox{$>$\kern-.75em\lower1ex\hbox{$\sim$}}}}
\def\ltwid{\mathrel{\raise.3ex\hbox{$<$\kern-.75em\lower1ex\hbox{$\sim$}}}}
\def\eq{eq.$\,$}
\def\eqs{eqs.$\,$}
\def\Eq{Eq.$\,$}

\singlespace
\rightline{astro-ph/9402037}
\rightline{CfPA--94--TH--11}
\rightline{UCSBTH--94--03}
\rightline{February 1994}

\vskip 3pt plus 0.3fill
\centerline{\titlefont Window Functions for CMB Experiments}

\vskip 3pt plus 0.2fill
\centerline{{\namefont Martin White}}

\vskip 3pt plus 0.1fill
\centerline{Center for Particle Astrophysics}
\centerline{and Departments of Astronomy and Physics,}
\centerline{University of California, Berkeley, CA 94720}

\vskip 3pt plus 0.2fill
\centerline{{\namefont Mark Srednicki}}

\vskip 3pt plus 0.1fill
\centerline{Department of Physics}
\centerline{University of California, Santa Barbara, CA 93106}

\vskip 3pt plus 0.3fill

{\narrower
\centerline{ABSTRACT}
\baselineskip=15pt
We discuss the applicability and derivation of window functions for cosmic
microwave background experiments on large and intermediate angular scales.
These window functions describe the response of the experiment to power in a
particular mode of the fluctuation spectrum.  We give general formulae,
illustrated with specific examples, for the most common observing strategies.}

\vskip 4pt

{\narrower
{\it Subject headings:} cosmic microwave background --- cosmology: theory}

\vskip 1in

\centerline{Submitted to {\it Monthly Notices}}

\vfill\eject
\baselineskip=12pt
\noindent 1. {\bf Introduction}
\vskip0.1in

It has become conventional in cosmic microwave background (CMB) studies
to describe the sensitivity of experiments by ``window'' or ``filter''
functions.  These functions describe the response of an experiment to
the power in a particular mode of the underlying fluctuation spectrum.
Plots of the window functions for various experiments are becoming
common (see e.g.~Bond~1990, Crittenden et al.~1993, Gorski~1993,
White, Scott \& Silk~1994),
and are increasingly used to compare different experiments.
We believe that it is useful to understand the derivation and meaning of
these window functions from a generic point of view.  Furthermore there are
experiments for which analysis by a window function is too complicated
to be useful, and it is important to understand what features of an experiment
lead to this situation.

We begin by assigning an ``ideal'' temperature $T(\nhat)$ to every point
on the sky; this is the temperature that would be measured by a perfect
experiment with an infinitely thin beam.  Here $\nhat$ is a unit vector which
can be expressed in the usual way in terms of the polar and azimuthal angles:
$$\nhat = (\sin\theta\cos\phi,\;
           \sin\theta\sin\phi,\;
           \cos\theta)\,.             \eqno(1)$$
We can expand the ideal temperature pattern in spherical harmonics:
$$ T(\nhat) = \sum_{\ell=1}^\infty\sum_{m=-\ell}^{\ell}
   a_{\ell m} Y_{\ell m}(\nhat)\,,      \eqno(2)$$
where we have removed the constant ($\ell=0$) term.
In {\it any} theory of fluctuations, the $a_{\ell m}$ are treated as
random variables; if we assume nothing more than rotational invariance,
we must have
$$ \bigl\langle a^{*}_{\ell m}a^{\phantom *}_{\ell' m'}\bigr\rangle_{\rm ens}
   \equiv C_\ell\,\delta_{\ell'\ell}\,\delta_{m'm}\,,       \eqno(3)$$
where the angle brackets denote an average over the statistical ensemble
of temperature fluctuations.  If the fluctuations are gaussian, all higher
point autocorrelation functions are given in terms of the two-point function,
and the set of $C_\ell$'s exhausts the content of the model.  Thus, any
gaussian model is completely specified by its predictions for the values
of the $C_\ell$'s, which will depend on some underlying parameters.
It is these parameters which we would like to measure with experimental data.
The computation of the $C_\ell$'s has been discussed by several authors,
including Peebles \& Yu~(1970); Wilson \& Silk~(1981); Bond \&
Efstathiou~(1984,1987); Vittorio \& Silk~(1984,1992); Holtzman~(1989);
Sugiyama \& Gouda~(1992); Dodelson \& Jubas~(1993); and Stompor~(1994).
In terms of the $C_\ell$'s, the two-point autocorrelation function of the
ideal temperatures $T(\nhat)$ is given by
$$\bigl\langle T(\nhat_1) T(\nhat_2) \bigr \rangle_{\rm ens}
 = {1\over 4\pi}\sum_{\ell=1}^\infty
   (2\ell+1)\,C_\ell\,\Pl(\nhat_1\cd\nhat_2)\,, \eqno(4)$$
where $\Pl(x)$ is a Legendre polynomial.

Of course no real experiment can measure the ideal temperatures.  In fact,
each experiment assigns a temperature, or a temperature ``difference'', to
points on the sky in a way which is generally unique to that experiment.
Call the temperature or temperature difference assigned to a point $\nhat$
by a particular experiment $\T(\nhat)$.
In general it will be linearly related to the ideal temperatures in some
neighborhood of $\nhat$:
$$\T(\nhat) = \int d\Omega_{\nhat'}\,M(\nhat,\nhat')\,T(\nhat')\,, \eqno(5)$$
where the {\it mapping function} $M(\nhat,\nhat')$ depends
on the detailed experimental strategy.  The mapping function is usually
too messy to compute in closed form, but it gives us a common way of
thinking about experiments, and as we will see it is closely related
to the more commonly used window function.

There are a number of different aspects of the experiment that go into
the mapping function.  One common to all experiments is the
{\it beam profile function} $B(\nhat,\nhat')$, which accounts for the
directional response of the antenna.  For large scale experiments, such as
COBE (Smoot et al.~1992) and FIRS (Ganga et al.~1993), which map the sky
this is the only effect, and the mapping function $M(\n,\n')$ is equal to
the beam profile function $B(\n,\n')$.

However, experiments on smaller scales are usually more complicated.
They typically use a chopping strategy,
possibly coupled with a smooth scan.  All cases are encompassed by
the following treatment.  We must specify the {\it beam position
function} $\nhat(t)$, or equivalently $\theta(t)$ and $\phi(t)$,
which tells us the position of the center of the beam
at time $t$.  We must also specify the {\it weighting} or {\it lock-in
function} $L(t)$, which tells us how different portions of the beam
trajectory are weighted in computing an experimental temperature,
and the overall {\it normalization} $N$.

Now consider a particular time interval, labelled by $i$, which runs
from $t=t_i-\half\delta_i$ to $t=t_i+\half\delta_i$.
To this time interval we assign an
average position $\nhatav_i$, given in terms of ${\overline\theta}_i$ and
${\overline\phi}_i$ by \eq(1), where
$$\eqalign{
{\overline\theta}_i &= {1\over\delta_i}\int_{t_i-\delta_i/2}^{t_i+\delta_i/2}
                        dt\,\theta(t)\,,\cr
\noalign{\medskip}
{\overline\phi}_i   &= {1\over\delta_i}\int_{t_i-\delta_i/2}^{t_i+\delta_i/2}
                        dt\,\phi(t)\,.\cr} \eqno(6)$$
To the average position $\nhatav_i$ we assign the temperature
$$\T(\nhatav_i)\equiv {N\over\delta_i}\int_{t_i-\delta_i/2}^{t_i+\delta_i/2}
                    dt\, L(t) \int d\Omega_{\nhat'} \,
                    B\bigl(\nhat(t),\nhat'\bigr) \,
                    T(\nhat')\,.\eqno(7)$$
\Eq(7) is completely general, and applies to all experiments.
It gives an implicit definition of the mapping function $M(\nhat,\nhat')$;
compare \eq(7) with \eq(5).
Below we will discuss a variety of possible choices for the beam position
function $\nhat(t)$ and the lock-in function $L(t)$.

To summarize, an experiment is completely specified by giving the
beam profile function $B(\nhat,\nhat')$, the beam position function
$\nhat(t)$, the lock-in function $L(t)$, and the normalization $N$.
All four must be given explicitly before the results of an experiment
can be analyzed or understood.  A measured value of ``$\Delta T$'' without
specification of all four of these experimental ingredients
cannot be interpreted.

Given the mapping function $M(\nhat,\nhat')$, we can compute the
{\it window function} $\Wl(\nhat,\nhat')$ as follows.
Let us begin with the two-point autocorrelation function
of the {\it experimental\/} temperatures:
$$\eqalignno{
\bigl\langle\T(\nhat_1)\T(\nhat_2)\bigr\rangle_{\rm ens}
 &= \int d\Omega_{\nhat'_1} \int d\Omega_{\nhat'_2}
    \, M(\nhat_1,\nhat'_1)
    \, M(\nhat_2,\nhat'_2)
    \, \bigl\langle T(\nhat'_1) T(\nhat'_2)\bigr\rangle_{\rm ens} \cr
 &= \int d\Omega_{\nhat'_1} \int d\Omega_{\nhat'_2}
    \, M(\nhat_1,\nhat'_1)
    \, M(\nhat_2,\nhat'_2)
    \, {1\over 4\pi}\sum_{\ell=1}^\infty
       (2\ell+1)\,C_\ell\,\Pl(\nhat'_1\cd\nhat'_2) \cr
 &= {1\over 4\pi}\sum_{\ell=1}^\infty
    (2\ell+1)\,C_\ell\,W_\ell(\nhat_1,\nhat_2)\,, & (8)\cr}$$
where the last equation defines the window function:
$$\Wl(\nhat_1,\nhat_2)
 \equiv  \int d\Omega_{\nhat'_1} \int d\Omega_{\nhat'_2}
         \, M(\nhat_1,\nhat'_1)
         \, M(\nhat_2,\nhat'_2)
         \, \Pl(\nhat_1'\cd\nhat_2')\,.   \eqno(9)$$
Often the window function is plotted in the literature as a function of
$\ell$ only.  This case corresponds to the window function at zero-lag,
$\Wl(\nhat,\nhat)$.  This is usually
independent of the choice of $\nhat$ to a {\it very} good approximation.
To specify slightly, we will assume that
$\Wl(\nhat,\nhat)$ is indeed independent of $\nhat$, and use the shorthand
notation $\Wl\equiv\Wl(\nhat,\nhat)$.  We will, however, also be interested
in the complete window function $\Wl(\nhat_1,\nhat_2)$.

\vskip\parskip
\noindent 2. {\bf Simple Window Functions}
\vskip0.1in

\noindent 2.1 {\it The beam profile}

As already noted, the simplest example of a window function is that which
arises when the finite size of the beam is the only effect, as is the case for
COBE (Smoot et al.~1992, Wright et al.~1994) and FIRS (Ganga et al.~1993).
In this case, the mapping function $M(\n,\n')$ is equal to the beam width
function $B(\n,\n')$.  If the beam profile is isotropic, then $B(\n,\n')$ is
a function of $\n\cd\n'$ alone, and we specialize to this case from here on.
We can expand $B(\n,\n')$ in Legendre polynomials:
$$ B(\nhat,\nhat')
   = {1\over4\pi}\sum_{\ell=0}^\infty (2\ell+1)\,B_\ell\,\Pl(\n\cd\n')\;,
							  \eqno(10)$$
and rewrite the $P_\ell$, using the addition theorem for spherical harmonics,
in \eq(9) to find that the window function is simply
$$ W_\ell(\n,\n') = \sum_{\ell=0}^\infty B^2_\ell\,\Pl(\n\cd\n')\;.
							  \eqno(11)$$
What we call $B_\ell^2$ is called $G_\ell$ by Wright et al.~(1994).
For a gaussian beam profile,
$$ B(\nhat,\nhat') ={1\over2\pi\sigma^2}\exp\bigl[-\theta^2/2\sigma^2\bigr]\;,
							  \eqno(12)$$
where $\theta\equiv\cos^{-1}(\nhat\cd\nhat')$, we have to a very good
approximation (Silk \& Wilson~1980, Bond \& Efstathiou~1984, White~1992)
$$B_\ell(\sigma)
= \exp\bigl[-\half\ell(\ell+1)\sigma^2\bigr]\;. \eqno(13)$$
In general the effect of the finite beam width is to provide a
high-$\ell$ cutoff at scales of the beam size $\ell\sim\sigma^{-1}$.
We note in passing that uncertainties in the value of $\sigma$
for a gaussian beam profile, or more generally the shape of
the beam profile, can result in significant uncertainties in
comparing experiment with theory, especially if the high-$\ell$
cutoff is in a range where $C_\ell$ is changing rapidly with~$\ell$.

\noindent 2.2 {\it Constant elevation scans}

We now turn to small scale experiments which use nontrivial beam position
functions $\n(t)$ and lock-in functions $L(t)$.
For these, it is possible to significantly simplify \eqs(7) and~(9) only if
the scan is performed at a constant $\theta$.
(Note that this needs to be the case only for one particular choice of
coordinates, which need not be equivalent to any of the usual choices.)
In this case the complete window function $\Wl(\n,\n')$ is a function only
of $|\phi-\phi'|$, an enormous simplification.
We specialize to this case for now, and will return to discuss the general
case later.

We first consider a stepped (as opposed to smooth) scan.  In this case the
beam is centered at a particular point $\overline\n = (\theta_0,\phi_0)$ on
the sky, and then ``chopped'' back and forth in the $\phi$-direction.
The instantaneous beam position $\n(t)$ is given by
$$\eqalign{
\theta(t) &= \theta_0 \;, \cr
  \phi(t) &= \phi_0 + \alpha_0\sin(\omega_{\rm c}t) \;, \cr}
							\eqno(14)$$
where $\alpha_0$ is half of the peak-to-peak chop angle, and
$\omega_{\rm c}/2\pi$ is the chop frequency.  (In practice it is a
few Hertz, but the window function turns out to be independent of
$\omega_{\rm c}$.)  Note that the angular separation on the sky is
measured by $\phi\sin\theta_0$.

In the case of a smooth scan, the beam is swept smoothly, at an
angular velocity of $\omega_{\rm s}$, in addition to being chopped.
The instantaneous beam position $\n(t)$ is now given by
$$\eqalign{
\theta(t) &= \theta_0 \;, \cr
  \phi(t) &= \phi_0 + \omega_{\rm s}t
             + \alpha_0\sin(\omega_{\rm c}t) \;, \cr} \eqno(15)$$
and the data must be binned, as in \eq(7), by integrating $t$ over the duration
time $\delta$ of a bin.  It practice $\delta$ is always a multiple of the
period of the chop; that is, $\omega_{\rm c}\delta/2\pi$ is an integer.

We are now in a position to compute the window function, assuming either
\eq(14) or \eq(15) for the instantaneous beam position.  We further assume
(again always the case in practice) that the lock-in function $L(t)$ has the
same periodicity as the chopping function, and that
$\omega_{\rm s}\ll\omega_{\rm c}$.  Again making use of the addition
theorem for spherical harmonics, we ultimately find from \eqs(7) and~(9) that
$$ W_\ell(\phi) = N^2 B^2_\ell(\sigma) \,
               {4\pi\over 2\ell+1}
               \sum_{m=-\ell}^\ell
               \left|Y_{\ell m}(\theta,0)\right|^2 \,
               L^2_m(\alpha_0) \,
               S^2_m(\Delta\phi) \,
               \cos(m\phi) \;,       \eqno(16) $$
where, up to an irrelevant phase,
$$L_m(\alpha_0) \equiv {\omega_{\rm c}\over2\pi}
\int_{-\pi/\omega_{\rm c}}^{+\pi/\omega_{\rm c}}
dt\,L(t)\,e^{im\alpha_0\sin(\omega_{\rm c}t)}    \eqno(17)$$
and
$$ S_m(\Delta\phi) \equiv j_0(m\Delta\phi/2) =
{\sin(m\Delta\phi/2) \over m\Delta\phi/2}        \eqno(18)$$
for a smooth scan.  Here $\Delta\phi = \omega_{\rm s}\delta$
is the size of the bins in $\phi$.  For a stepped scan, $S_m=1$.
For more details in the context of specific choices of $L(t)$, see
Dodelson \& Jubas~(1993); White et al.~(1993).

If we can neglect the curvature of the line and assume that it is an arc of
a great circle (usually a very good approximation), then we can set
$\theta=\pi/2$ and let $\phi$ be the angle on the sky; in this case
$$  W_\ell(\phi)  = N^2 B^2_\ell(\sigma) \sum_{r=0}^\ell
{(2\ell-2r)!(2r)!\over[2^\ell r!(\ell-r)!]^2}\,L^2_{\ell-2r}(\alpha_0)\,
S^2_{\ell-2r}(\Delta\phi)\,\cos[(\ell-2r)\phi] \eqno(19) $$
which is easy to implement numerically.  Note that now $\phi$,
$\Delta\phi$, and $\alpha_0$ are all defined as angles {\it on the sky}.

\noindent 2.3 {\it The lock-in}

The simplest lock-in function is that for a ``square
wave chop'' recently used by Tenerife (Hancock et al.~1994), MSAM
(Cheng et al.~1994), OVRO (Myers, Readhead \& Lawrence~1993) and Python
(Dragovan et al.~1994).
In this strategy the temperature assigned to $\phi_0$ is a weighted sum of
temperatures along a line, which we assume to be of constant elevation.
The telescope moves rapidly between the observed points, stopping and taking
data at set positions on the line.
The weights assigned to points on the sky for three different ``switching
strategies'' are
\vskip12pt
\centerline{
\vbox{ \offinterlineskip
\halign {#\hfil\vrule& \hfil#\hfil& \hfil#\hfil&
\hfil#\hfil& \hfil#\hfil& \hfil#\hfil\cr
\qquad&\qquad&\qquad&\qquad&\qquad&\qquad \cr
$(\phi-\phi_0)/\alpha_0$\ &$-1$&$-\third$&$0$&$+\third$&$+1$\cr
\noalign{\hrule}
\strut&\omit&\omit&\omit&\omit&\omit\cr
2-beam&$+1$&\omit&\omit&\omit&$-1$\cr
\strut&\omit&\omit&\omit&\omit&\omit\cr
3-beam&$-\half$&\omit&$+1$&\omit&$-\half$ \cr
\strut&\omit&\omit&\omit&\omit&\omit\cr
4-beam&${\textstyle +{1\over 4}}$&${\textstyle -{3\over 4}}$&\omit&
${\textstyle +{3\over 4}}$&${\textstyle -{1\over 4}}$\cr }} }
\vskip12pt \noindent
In our notation such a strategy is implemented by taking $L(t)$ to be a linear
combination of Dirac delta functions.
For a two-beam, three-beam, or four-beam switching strategy, we have
$$ {\omega_{\rm c}\over 2\pi} L(t) = \cases{
\vphantom{i}
+ \delta(t-t_{\rm c})
- \delta(t+t_{\rm c})               &(2-beam)\cr
\noalign{\medskip}
\vphantom{i}
- \frac1/2 \delta(t-t_{\rm c})
+ \delta(t)
- \frac1/2 \delta(t+t_{\rm c})      &(3-beam)\cr
\noalign{\medskip}
\vphantom{i}
+ \frac1/4 \delta(t-t_{\rm c})
- \frac3/4 \delta(t-\xi t_{\rm c})
+ \frac3/4 \delta(t+\xi t_{\rm c})
- \frac1/4 \delta(t+t_{\rm c})      &(4-beam)\cr} \eqno(20)$$
where $\xi=\sin^{-1}(1/3)$ and $t_{\rm c}=\pi/2\omega_{\rm c}$ is the time
to chop from $\phi = \phi_0$ to $\phi = \phi_0+\alpha_0$.
In this case the mapping function $M(\nhat,\nhat')$ defined through \eq(7)
reduces to a weighted sum of beam profile functions $B(\nhat,\nhat')$.  From
\eq(17) we find immediately that, up to an irrelevant phase,
$$L_m(\alpha_0) = \cases{
2\sin(m\alpha_0)                    &(2-beam)\cr
\noalign{\medskip}
2\sin^2(m\alpha_0/2)                &(3-beam)\cr
\noalign{\medskip}
\frac1/2[\sin(m\alpha_0) - 3\sin(\frac1/3 m\alpha_0)]
				    &(4-beam)\cr} \eqno(21)$$
Notice that $L_m(\alpha_0)$ scales as $\alpha_0^{n-1}$ for the $n$-beam
switching strategy.  In general the window function for any kind of
differencing experiment is suppressed at low $\ell$, since any long wavelength
perturbation is removed by the differencing.
Since the low $\ell$ cutoff is controlled by $\alpha_0$ while the high $\ell$
cutoff is specified by $\sigma$, one can increase both the height and width of
$\Wl$ by separating these scales as much as possible.

To make contact with forms of $\Wl$ frequently quoted in the literature,
we note that we can substitute \eq(21) into \eq(19) with $\phi=0$ (zero lag)
and use the addition theorem for spherical harmonics to obtain
$$\Wl = B_\ell^2(\sigma) \cases{
2\bigl[ 1-\Pl(\cos2\alpha_0) \bigr] &(2-beam) \cr
\noalign{\medskip}
\frac1/2\bigl[3-4\Pl(\cos\alpha_0)+\Pl(\cos2\alpha_0)\bigr] &(3-beam)\cr
\noalign{\medskip}
\frac1/8\bigl[10-15\Pl(\cos\frac2/3\alpha_0)+6\Pl(\cos\frac4/3\alpha_0)
-\Pl(\cos2\alpha_0)\bigr] &(4-beam)\cr}           \eqno(22)$$

Three other illustrative choices of the lock-in function for differencing
experiments are the ``square-wave lock-in,''
$$\eqalign{
L(t) &= 2\mathop{\hbox{sgn}}[\alpha_0\sin(\omega_{\rm c}t)]\;,\cr
L_m(\alpha_0) &= 2\,H_0(m\alpha_0)\;,\cr}         \eqno(23)$$
used by SP91 (Gaier et al.~1991) \& ARGO (de Bernardis et al.~1994),
the ``sine-wave lock-in,''
$$\eqalign{
L(t) &= \pi\sin(\omega_{\rm c}t)\;,\cr
L_m(\alpha_0) &= \pi\,J_1(m\alpha_0)\;,\cr}       \eqno(24)$$
used by MAX (Gundersen et al.~1993, Meinhold et al.~1993)
and the double-angle ``cosine lock-in'',
$$\eqalign{
L(t) &= \pi\cos(2\omega_{\rm c}t)\;,\cr
L_m(\alpha_0) &= \pi\,J_2(m\alpha_0)\;,\cr}       \eqno(25)$$
used by Saskatoon (Wollack et al.~1994).
Here $H_0(x)$ is the Struve function, and $J_n(x)$ are Bessel
functions of the first kind.  The numerical prefactors are chosen
so that
$${\omega_{\rm c}\over2\pi} \int_{-\pi/\omega_{\rm c}}^{+\pi/\omega_{\rm c}}
 dt\,\bigl|L(t)\bigr| = 2 \;,     \eqno(26)$$
which is a common way of normalizing an experiment (more on this below).
Sometimes the $L_m(\alpha_0)$'s of \eqs(23) and~(24) are approximated
by the $L_m(\alpha_0)$ for a ``2-beam chop'', as given in \eq(20),
but this can result in significant errors.  For example, after
taking into account the normalizations used by these experiments, we
find that for SP91 the approximation is off by $20\%$
(see Dodelson \& Stebbins~1994) while for MAX it differs by $10\%$.
Also, the Saskatoon lock-in function of \eq(25) is only roughly
approximated by the 3-beam result of \eq(21).

\noindent 2.4 {\it The normalization}

Finally we consider the normalization factor $N$.  For SP91 and MAX this is
chosen so that if there is a sharp boundary between two regions of constant
temperatures $T_1$ and $T_2$, then aiming at a point on the boundary,
and chopping perpendicular to the boundary, gives $\T=T_2-T_1$.
For SP91, the normalization is computed assuming a perfect, point-like beam,
corresponding to $\sigma=0$ in \eq(13).
For a perfect beam, the lock-in factor $L_m(\alpha_0)$ given in \eqs(21--25)
is already normalized so that $N=1$.  However, if the normalization is done
assuming the actual beam profile of the experiment (as is the case for MAX
and Saskatoon), then there are corrections which must be computed.
The result for MAX was presented in Srednicki et al.~(1993), but here we
give a more general treatment.

We first make the ``flat sky approximation'' near the $T_1$--$T_2$
boundary, which we shall take to be the line of longitude $\phi=0$.
We treat $x\equiv\phi$ and $y\equiv\frac\pi/2-\theta$ as cartesian coordinates.
For $\n=(0,0)$ and $\n'=(x,y)$, \eq(12) for the beam profile becomes
$$B(x,y)={1\over2\pi\sigma^2}\,e^{-(x^2+y^2)/2\sigma^2}\;.   \eqno(27)$$
\Eq(5) can now be written as
$$\T(0,0) = \int^{+\infty}_{-\infty}dx\,dy\,M(x,y)T(x,y)\;,  \eqno(28)$$
where the mapping function is given by
$$M(x,y) = {N\omega_{\rm c}\over(2\pi)^2\sigma^2}
\int_{-\pi/\omega_{\rm c}}^{+\pi/\omega_{\rm c}} dt\,L(t)
\exp\left(-{[x-\alpha_0\sin(\omega_{\rm c}t)]^2+y^2\over2\sigma^2}\right)\;.
                                                              \eqno(29) $$
Take the temperature profile to be $T(x,y)=T_0\,\theta(x)$, where $\theta(x)$
is the step function, and demand that $\T(0,0) = T_0$.
For the MAX lock-in function given by \eq(24), we get
$$\eqalign{
1 &= \frac1/4 N\int_{-\pi}^{+\pi}dr \sin r\mathop{\rm erf}(\gamma\sin r)\cr
\noalign{\medskip}
  &= N\left(\half\gamma\sqrt\pi\,\right){}_1F_1(\frac1/2,2;-\gamma^2)\;,\cr}
                                                      \eqno(30) $$
where $\gamma=\alpha_0/\sqrt2\sigma$, erf is the error function, and ${}_1F_1$
is the confluent hypergeometric function.  For MAX, with $\alpha_0=0\degrees65$
and $\sigma=0.425\times 0\degrees5$, this gives $N^2=1.13$.
(Note that it is $N^2$ which appears in the window function.)

For Saskatoon, the only change is that the normalizing temperature profile
is $T_0$ where $M(x,y)\ge0$, and zero elsewhere.  The mapping function
follows from substituting \eq(25) into \eq(29), which also implicitly defines
the region where $T(x,y)\ne0$ in \eq(28).  Setting $\T(0,0)$ to $T_0$ as
before gives
$$1 = {N\over 2\sigma\sqrt{2\pi}}\int dx\ \int_{-\pi}^{\pi} dr \cos(2r)
\exp\left(-{[x-\alpha_0\sin r]^2\over 2\sigma^2}\right)\;.\eqno(31)$$
where the integration is over the region of $x$ where $M(x,y)\ge0$.
With $\alpha_0=2\degrees45$ and $\sigma=0.425\times 1\degrees44$, we find
$N^2=1.74$, quite a large correction.  Also note that the mapping $M(x,y)$
defined by \eq(25) and \eq(29) is not well approximated by a square wave chop
pattern, which would be three gaussian beam profiles at $x=-\alpha_0$, $0$
and $\alpha_0$ with weights $-\half$, $+1$ and $-\half$ respectively.

\vskip\parskip
\noindent 3. {\bf Other Strategies}
\vskip0.1in

One other ``differencing'' strategy that has been proposed recently is used
by the ``White Dish'' experiment (Tucker et al.~1993), which assigns to the
point $\nhat$ a temperature $\T(\nhat)$ which is given by a particular
weighted average of measured temperatures in a circle around that point.
The actual strategy used is {\it very\/} difficult to model effectively.
However, if we modify their ``Method II'' analysis to neglect binning, then it
is straightforward to compute $\Wl=\Wl(\nhat,\nhat)$.  While the off-diagonal
elements cannot be simply constructed, a numerical procedure similar to that
used in Srednicki et al.~(1993) would be feasible.

To get $\Wl$, we first rotate coordinates so that $\nhat$ is at $\theta=0$
with the circle being in $\phi$ at fixed $\theta=\theta_0$.
The analogue of \eq(7) is now to extract the $n$th harmonic of the
temperature around the circle, which corresponds to a window function
of the form [c.f.~\eq(16)]
$$\eqalign{ \Wl &= {4\pi\over 2\ell+1} N^2 B^2_\ell(\sigma) {\pi^2\over 2}
   \left| Y_{\ell n}(\theta_0,0)\right|^2 \cr
\noalign{\medskip}
  &\simeq {\pi^2\over 2} N^2 B^2_\ell(\sigma)
    \left({(\ell+n)!\over (\ell-n)!\ell^{2n}}\right)
   J_n^2(\ell\theta_0)\cr } \eqno(32) $$
where in the last line the limit $\theta_0\ll 1$ has been used.  The term
in parenthesis on the last line is very close to unity for $\ell\gg n$.
Assuming that a temperature profile which is $T_0$ for both $0\le\phi<\pi/2$
and $\pi\le\phi<3\pi/2$ and zero elsewhere would be assigned a temperature
difference of $\T=T_0$, when $n=2$, we get $N=1$.
A less artificial normalization would require more information than is
specified in the paper (Tucker et al.~1993).
The White Dish experiment has $\theta_0=14'$, $\sigma=0.425\times 12'$, and
$n=2$, with the resulting ``temperatures'' binned into 4 positions
$T_1$,...,$T_4$ in a square of side $\theta'=23.6'$.
The four temperatures are assigned to consecutive corners
clockwise around the square (Tucker et al~1993).

\Eq(32) can be compared with the window function for a ``square wave chop''
procedure where we simply sum the temperatures in a square:
$\T=\half(T_1-T_2+T_3-T_4)$, neglecting how they were measured, to obtain
$$ \Wl = B^2_\ell(\sigma)
\left[ 1-2P_\ell(\cos\theta')+P_\ell(\cos\sqrt{2}\theta')\right] $$
Assuming that $N=1$, the two window functions peak in the same place
($\ell\sim500$), but differ by $20\%$ at the peak and have a different
scaling with $\ell$ off the peak.  Additionally, neither of these methods
accurately reflects the correlations induced by coarse binning of the data.

With an analysis procedure as difficult to model as White Dish, the window
function approach is of limited utility and one should resort to Monte-Carlo
simulations of the observing strategy, for each theory being tested.
Alternatively the applicability of a window functions should be kept in mind
when the analysis procedure is designed.

\vskip\parskip
\noindent 4. {\bf Scans at Varying Elevation}
\vskip0.1in
If data points are not taken at constant elevation, such as in the GUM
scan of MAX (Gundersen et al.~1993), and chopping is used, computing
the window function at nonzero lag ($\n\ne\n'$) is impossible analytically.
The information which is needed to compare data with a theory is the
autocorrelation function of the experimental temperatures, \eq(8),
computed with the $C_\ell$'s of the theory in question.
It is then usually easier to compute this directly, using numerical methods,
than to compute the complete window function $W_\ell(\n,\n')$.  However, even
numerical methods become cumbersome if the observing strategy is complex.
Accounting for chopping generally requires that a double integral be done
numerically (Srednicki et al.~1993).
If there is, in addition, a smooth (as opposed to stepped) scan, then a
quadruple integral must be done numerically, and this is not feasible in
general.
If the data is binned finely enough, then the effects of the smooth scan are
small, and this is not a problem.
Here ``finely enough'' means $\Delta\phi\ltwid\alpha_0$, fortunately the case
for MAX-GUM.

\vskip\parskip
\noindent 5. {\bf Conclusions}
\vskip0.1in

Using the formulae presented in this letter it is possible to compute
window functions for most of the current large and intermediate scale
CMB anisotropy experiments.  We stress however that there exist several
generic cases in which the window function approach is not the optimal
method of analysis.  These are when the correlation matrix
$\bigl\langle\T(\nhat_1)\T(\nhat_2)\bigr\rangle_{\rm ens}$
is anisotropic (such as in the case of the GUM scan of the MAX experiment or
multiple scans of the SP91 experiment) or if the experimental procedure makes
analytic calculation of the window function difficult.  In these cases,
it is generally much easier to compute \eq(8) directly (by numerical methods),
or to do a Monte Carlo analysis of the experiment, than it is to try to
calculate the window function $W_\ell(\nhat,\nhat')$.

Both diagonal and off-diagonal window functions are easy to construct for
experiments in which the scanning and chopping directions are constant,
regardless of the type of chopping (square wave, sine, cosine), and regardless
of whether the scan is smooth or stepped.  For these cases, the window
function provides an efficient method for comparing theory with data.
However, with the current refined state of CMB anisotropy measurements,
it is important to used the {\it right} window function.  A simple
stepped-scan, square-wave chop approximation to all experiments is no
longer accurate enough for the quality of data now available.

\medskip
We would like to thank Douglas Scott for useful conversations and comments
on the manuscript.
This work was supported in part by NSF Grant Nos.~PHY--91--16964
and AST--91--20005.  M.W.~acknowledges the support of a fellowship
from the TNRLC.


\vskip\parskip
\noindent {\bf References}
\vskip0.1in
\frenchspacing
\parindent=0truept
\baselineskip=12pt

\abook de Bernardis, P., et al., 1994;ApJ;to appear

\abook Bond, J. R., 1990; {\rm in} ``Frontiers in Physics: From Colliders to
Cosmology: Proceedings of the Fourth Lake Louise Winter Institute'';
ed.~A. Astbury et al., p.182 (Singapore, World Scientific)

\aref Bond, J. R. \& Efstathiou, G., 1984;ApJ;285;L45

\aref Bond, J. R. \& Efstathiou, G., 1987;MNRAS;226;655

\aref Bond, J. R., et al., 1991;Phys. Rev. Lett.;66;2179

\abook Bunn, E., White, M., Srednicki, M.~\& Scott, D., 1994;ApJ;to appear

\abook Cheng, E. S., et al., 1994;ApJ;to appear

\aref Crittenden, R., et al., 1993;Phys. Rev. Lett.;71;324

\abook Dragovan, M., et al., 1994;ApJ;submitted

\aref Dodelson, S. \& Jubas, J. M., 1993;Phys. Rev. Lett.;70;2224

\abook Dodelson, S. \& Stebbins, A., 1994;ApJ;submitted

\aref Gaier, T., et al., 1992;ApJ;398;L1

\aref Ganga, K., et al., 1993;ApJ;410;L57

\aref Gorski, K. M., 1993;ApJ;410;L65

\aref Gundersen, J. O., et al., 1993;ApJ;413;L1

\aref Hancock, S., et al., 1994;Nature;367;333

\aref Holtzman, J. A., 1989; ApJ (Supp);71;1

\aref Meinhold, P. R., et al., 1993;ApJ;409;L1

\aref Myers, S. T., Readhead, A. C. \& Lawrence, C. R., 1993;ApJ;405;8

\aref Peebles, P. J. E. \& Yu, J. T., 1970;ApJ;162;815

\aref Readhead, A. C. S. \& Lawrence, C. R., 1992;Annu. Rev. Astron. \&
Astrophys.;30;653

\aref Silk, J. \& Wilson, M. L., 1980;Physica Scripta;21;708

\aref Srednicki, M., White, M., Scott, D.~\& Bunn, E.,
1993;Phys Rev Lett;71;3747

\abook Stompor, R., 1994;Astron. Astrophys.;in press

\aref Sugiyama, N. \& Gouda, N., 1992;Prog. Theor. Phys.;88;803

\aref Tucker, G. S., et al., 1993;ApJ;419;L45

\aref Vittorio, N. \& Silk, J., 1984;ApJ;285;L39

\aref Vittorio, N. \& Silk, J., 1992;ApJ;385;L9

\aref White, M., 1992;Phys. Rev.;D42;4198

\aref White, M., Krauss, L., \& Silk, J., 1993;ApJ;418;535

\abook White, M., Scott, D., \& Silk, J., 1994;
Annu. Rev. Astron. \& Astrophys.;to appear

\aref Wilson, M. L. \& Silk, J., 1981;ApJ;243;14

\abook Wollack, E. J., et al., 1994;ApJ;submitted

\arep Wright, E. L., et al., 1994;ApJ;submitted

\nonfrenchspacing

\end